\documentclass[conference,10pt]{IEEEtran}
\usepackage{cite}
\usepackage{amsmath,amssymb,amsfonts}
\usepackage{algorithmic}
\usepackage{graphicx}
\usepackage{textcomp}
\usepackage{xcolor}
\usepackage{hyperref}
\usepackage{makecell} 
\usepackage{todonotes}
\usepackage{color}
\usepackage{latexsym}
\usepackage{times}
\usepackage{amssymb}
\usepackage{latexsym}
\usepackage{amsmath}
\usepackage{algorithm}
\usepackage{wrapfig,epsf}
\usepackage{amsthm}
\usepackage{float}
\usepackage{floatflt}  
\usepackage{moreverb}
\usepackage{color,soul}
\usepackage{latexsym}
\usepackage{graphicx,subfigure}
\usepackage{times}

\usepackage{tikz}
\usepackage{array}
\usepackage{colortbl}

\newcommand{\yes}{\tikz\draw[fill=black,draw=black] (0,0) circle (0.5ex);} 
\newcommand{\no}{\tikz\draw[fill=white,draw=black] (0,0) circle (0.5ex);} 
\newcommand{\partialcircle}{%
  \tikz{%
    \draw[draw=black] (0,0) circle (0.5ex);
    \begin{scope}
      \clip (0,0) circle (0.5ex);
      \fill[black] (-0.5ex,0) rectangle (0, 0.5ex);
      \fill[black] (-0.5ex,0) rectangle (0, -0.5ex);
    \end{scope}
  }
}

\def\BibTeX{{\rm B\kern-.05em{\sc i\kern-.025em b}\kern-.08em
    T\kern-.1667em\lower.7ex\hbox{E}\kern-.125emX}}

\begin{document}

\title{Towards Privacy-Preserving Revocation of Verifiable Credentials with Time-Flexibility}

\author{\IEEEauthorblockN{1\textsuperscript{st} Francesco Buccafurri}
\IEEEauthorblockA{
\textit{University Mediterranea of Reggio Calabria}\\
Reggio Calabria, Italy \\
bucca@unirc.it}
\and
\IEEEauthorblockN{2\textsuperscript{nd} Carmen Licciardi}
\IEEEauthorblockA{
\textit{University Mediterranea of Reggio Calabria}\\
Reggio Calabria, Italy \\
carmen.licciardi@unirc.it}

}

\maketitle
\begin{abstract}
Self-Sovereign Identity (SSI) is an emerging paradigm for authentication and credential presentation that aims to give users control over their data and prevent any kind of tracking by (even trusted) third parties. In the European Union, the EUDI Digital Identity wallet is about to become a concrete implementation of this paradigm. However, a debate is still ongoing, partially reflecting some aspects that are not yet consolidated in the scientific state of the art. Among these, an effective, efficient, and privacy-preserving implementation of verifiable credential revocation remains a subject of discussion. In this work-in-progress paper, we propose the basis of a novel method that customizes the use of anonymous hierarchical identity-based encryption to restrict the Verifier access to the temporal authorizations granted by the Holder. This way, the Issuer cannot track the Holder's credential presentations, and the Verifier cannot check revocation information beyond what is permitted by the Holder.
\end{abstract}

\section{Introduction and Related Work}

Self-Sovereign Identity (SSI) \cite{allen2016ssi,preukschat2021self} has emerged as a pivotal paradigm in the evolving landscape of digital identity, offering a decentralized and user-centric approach that aims to increase the privacy of existing digital identity approaches
while (at least) preserving their security and
interoperability.

Although the trustworthiness of Verifiers and Issuers can be assessed and managed through trust models and cryptographic mechanisms
\cite{buccafurri2024can,de2023multi}, open challenges remain regarding the risk of honest-but-curious behavior or even more adversarial threats.
Among these, designing a robust and practical method for verifiable-credential revocation that ensures the impossibility for both Issuer and Verifier to track information regarding the Holder is still an open problem \cite{freitag2022revocation,roio2024sdbls}.
This debate is also alive within the standardization process of the European Digital Identity wallet (EUDI) \cite{eudi2021}, as witnessed by recent
Internet-Drafts \cite{demarco2024oauth,IETF2025}.
The goal is to prevent the Verifier from monitoring the status of a Verifiable Credential (VC) over time, while ensuring that the use of the VC remains unknown to all parties—including the Issuer—except for the Holder and the Verifier.
Apart from the above references, some recent scientific papers related to this problem are available in the literature
\cite{manimaran2024prevoke,hoops2025crset,roio2024sd}.
However, they either use blockchain \cite{roio2024sd} (which is little practical and very expensive) or design complex systems
\cite{manimaran2024prevoke,hoops2025crset} that significantly limit their applicability in real-life implementations of SSI, such as the EUDI wallet.
Another recent proposal is \cite{mazzocca2024evoke}. It relies on distributed ledger technology, focuses on IoT, and considers a threat model different from that of this paper.
Notably, the proposal given in \cite{demarco2024oauth} (which, as an Internet-Draft, is inherently designed for practical use), faces a trade-off between high bandwidth consumption Holder-side (due to the periodical status assertion reception from the Issuer)
and the freshness of revocation information.

In this work-in-progress paper, we propose the basis of a novel approach that customizes the use of anonymous hierarchical identity-based encryption to 
achieve the stated privacy goals stated while overcoming the limitations of
\cite{demarco2024oauth} and \cite{IETF2025}. 
Within a significant threat model, we conduct a preliminary security assessment of our approach through an attack-based analysis.
Interestingly,
our proposal includes a very novel feature.
The access of the Verifier to revocation information can be restricted not only to the moment of VC presentation but also to specific time frames authorized by the Holder, thus enabling \textit{time-flexibility}. This approach effectively addresses many practical use cases, such as a competition where applicants include a VC in their documentation, and its status needs to be verified only at a later stage during the committee’s evaluation process.

The paper is structured as follows. In Section \ref{sec:background}, we provide some background notions about anonymous hierarchical identity-based encryption.
In Section \ref{sec:threat_model}, we introduce our threat model. The motivations behind our work are discussed in Section \ref{sec:motivations}, where we present a comparative analysis of achieved features by examining existing approaches and our method. It is described in Section \ref{sec:solution}, which also demonstrates how our solution meets the features outlined in Section \ref{sec:motivations}.
The security analysis is sketched in Section \ref{sect:security}.
We draw our conclusions and discuss future work in Section \ref{sec:conclusion}.

\section{Background} \label{sec:background}
Our paper focuses on Self-Sovereign Identity and proposes a solution for privacy-preserving revocation of verifiable credentials based on Anonymous Hierarchical Identity-Based Encryption.
We assume the reader is familiar with the SSI paradigm. Therefore, in this section, we give some basic notions about Anonymous Hierarchical Identity-Based Encryption.

Anonymous Hierarchical Identity-Based Encryption (AHIBE) is an advanced cryptographic scheme that extends the concept of Identity-Based Encryption (IBE) \cite{boneh2001identity} to hierarchical structures while providing anonymity for recipients. IBE allows public keys to be derived from user identities, eliminating the need for a traditional public key infrastructure (PKI). In a hierarchical setup, the root authority can delegate key generation capabilities to subordinate authorities, enabling efficient and scalable key management.
In Hierarchical Identity-Based Encryption (HIBE), subordinate identities can be represented as strings obtained by concatenating the root identity with a string associated with the subordinate. This concatenation naturally reflects the hierarchical relationship between different levels of the identity structure.
Formally, if ${ID}_{{r}}$ represents the root identity and ${ID}_{{s}}$ represents the identity of a subordinate, the subordinate's identity can be expressed as: ${ID}_{{s}} = {ID}_{{r}} \parallel {ID}_{{s}}$
where $\parallel$ denotes string concatenation. For multiple levels of hierarchy, this process can be iteratively applied.

In Identity-Based Encryption (IBE) and its hierarchical variant, the presence of a Private Key Generator (PKG) is fundamental. The PKG is a trusted authority responsible for generating and distributing private keys to users based on their identities. 
The PKG first generates a master public key and a corresponding master secret key. The master public key is made publicly available, enabling anyone to encrypt messages for a user by using the recipient's identity as the public key. The master secret key, however, is kept private and is used by the PKG to derive user-specific private keys.
In Hierarchical Identity-Based Encryption (HIBE), the PKG can delegate key generation to subordinate PKGs, which operate at lower levels of the hierarchy. This delegation process enhances scalability while maintaining the security of the system. The trust placed in the PKG is crucial since it has access to the master secret key, making it a potential point of compromise. To mitigate this risk, techniques such as distributed PKGs or threshold cryptography are often employed.
The anonymity property in AHIBE ensures that cipher-texts do not reveal the identity of the recipient, even to an adversary with access to public parameters. This feature is crucial for privacy-preserving applications such as secure communications, access control, and cloud computing. AHIBE constructions typically rely on bilinear pair pairings. However, quantum-safe lattice-based Anonymous Hierarchical Identity-Based Encryption (AHIBE) is available such as \cite{van2023quantum}.

In our paper, AHIBE is used unconventionally. Specifically, we adopt AHIBE as a workaround to grant temporal encryption-enforced access control to ephemeral identities obtained by the concatenation of the Holder's identity (i.e., the root) and the time.

\section{Threat Model} \label{sec:threat_model}

As adversaries, we consider the Issuer, the Verifier, and the Holder.
The Verifier is honest but curious (i.e., semi-trusted). The Issuer acts as an honest-but-curious adversary but can also act dishonestly as a censor.
The Holder can be malicious.
The attacks we consider are: (\textbf{Attack 1}) the Verifier tries to track the revocation information of a given Verifiable credential (VC) over time (with respect to the time of the VC presentation or possible temporal authorizations granted by the Holder); (\textbf{Attack 2}) the Issuer tries to track the use of VCs (this should not possible in an SSI system); (\textbf{Attack 3}) the Issuer (acting as a censor) tries to inhibit a valid VC presentation; 
(\textbf{Attack 4}) the Holder tries to hide from the Verifier an revocation information regarding the presented VC.

\section{Motivations} \label{sec:motivations}

This paper aims is to propose a new method for VC revocation that overcomes the limitations of the existing ones.
To better explain the motivations of this goal, first, we identify the features about which a revocation method should be evaluated.
The features representing the dimensions of our analysis are:

\noindent
\textbf{Scalability}: It is the ability of the solution to handle an increasing number of verification requests
    and verifiable credentials without degrading performance or introducing bottlenecks.
    
\noindent
\textbf{Bandwidth consumption Verifier-side}: It is the amount of data that the Verifier must download or exchange to obtain and verify the status of VC. 

\noindent
\textbf{Bandwidth consumption Holder-side}: It is the amount of data that the Holder must download or transmit to verify and prove the revocation status of their credential.
    
\noindent
\textbf{Untraceability by the Verifier}: It means that the Verifier must not be able to track any state changes of the VC without the explicit consent of the Holder, expected at the time of VC presentation. Any tracking outside these circumstances is allowed only with the Holder's explicit authorization. 

\noindent
\textbf{Time flexibility}: When untreceability by the Verifier is achieved, we would like to allow the Holder to grant temporal authorizations to the Verifier in a flexible way, regarding both the past, the present, and the future.
    As an example, let us consider the case of a competition where applicants include VC in their documentation, and its status needs to be verified only at a later stage during the evaluation process by the committee. Similarly, in a forensic context, a law enforcement agency may need to retroactively verify the status of a VC for a specific past date.

\noindent
\textbf{Untraceability by the Issuer}: It means that the Issuer must not be able to track the use of a VC by the Holder. It is important to highlight that even if the Holder is unaware of the identity of the Verifier with whom the VC is presented, significant privacy issues may still arise. For instance, consider the case of a healthcare VC that is preset on a monthly basis: an \textit{honest-but-curious} Issuer, leveraging background knowledge, could infer sensitive information, such as the fact that the Holder is undergoing a specific therapy.

\noindent
\textbf{Censorship resistance}: We would like to prevent an untrusted Issuer from inhibiting a particular holder from using a valid verifiable credential (VC) without such dishonest behavior being easily detected. 

\noindent
\textbf{Content-flexibility}: The status of VC should go beyond a simple binary classification (e.g., revoked/not revoked) to more accurately represent various conditions, such as suspension, conditional validity, or other relevant information. For instance, consider a driver's license point system. In practice, there are several states that provide more detailed information on its validity. A license can be: valid; revoked; suspended, when the Holder cannot use it temporarily; and conditioned, when the Holder can only drive under specific conditions.  

\noindent
\textbf{Freshness}: The verification process should include the detection of any recent status changes, ensuring an up-to-date and accurate analysis of the validity of the VC.

To analyze existing approaches, it is necessary to make a basic distinction about the default assumption adopted. 

Is the default that the VC has full validity if its status is not explicitly changed? If so, the Issuer provides only the information that restricts the validity of a VC. 
Otherwise, i.e., the default is that the VC has no validity until the exact status is stated. In this case, the Issuer provides the exact description of the VC status.

It is possible to observe that the size of the domain of VCs is related to the amount of information we consider in them. The domain size of the possible VC status descriptions turns out to be significantly larger since it includes all possible conditions in which a VC may be (e.g., valid, revoked, expired, suspended, etc.). In contrast, the domain size of restriction information only specifies the revocation (or restriction) of the validity of a VC; thus, the dimension is smaller. 
Consider the simplest on/off case of revocation. We expect that only a very small fraction of VCs will be revoked. Thus, it is conceptually much more expensive to materialize the of description the VC status instead of restriction information (i.e., the small subset of revocations).

At this point, on the basis of the existing solutions, we categorize the possible approaches
from a conceptual point of view:

\noindent
\textbf{Revocation Information List (RIL)}: Similarly to CRL used in X.509, the Issuer publishes the list of VC revocation information. This list is downloaded by the Verifier.

\noindent
\textbf{Online status Protocol (OSP)}: Similarly to OCSP used in X.509, the Verifier queries the Issuer to obtain the status of a specific VC.

\noindent
\textbf{VC Status List (VSL)}:  As described in \cite{oauth-status-assertions}
the Issuer publishes the list of status information about all the VCs. This list is downloaded by the Verifier.

\noindent
\textbf{Status Assertion (STA) }:  As in  \cite{demarco2024oauth}
the Issuer sends periodically to the Holders a signed object that demonstrates the validity status of a VC. The Holder can present this object together with VC.

\noindent
\textbf{On demand Status Assertion (OSA)}: A statement about the status of a VC is provided only when requested by the Holder, rather than being pre-distributed. Again, the Holder can present it together with the VC.

Now, we briefly analyze the approaches above in light of the features earlier identified.

\textbf{Scalability}:
\textbf{RIL} has limitations in terms of scalability because it requires storing and publishing a complete revocation information list. This can be inefficient when the number of revocation information to be managed is large. However, recall that the revocation information domain is a small subset of the verifiable credential set.
\textbf{OSP} requires a separate request for each verified certificate, which implies a higher load for each individual request and centralized management on the Issuer's server.
\textbf{VSL} has lower scalability compared to \textbf{RIL}, because explicit status information per released VC must be kept and published by the Issuer.
\textbf{STA} has increased issues of scalability from the computation point of view, because the Issuer must periodically send assertions to all the Holders.
\textbf{OSA} mitigates the scalability problems of \textbf{STA}, because assertions are sent only on demand, but its scalability is lolwer than \textbf{VSL}, because the Issuer
has to process requests. 

\textbf{Bandwidth consumption Verifier-side}.
\textbf{RIL} has a high bandwidth consumption for the Verifier, as it requires the complete download of the revocation information list, which is inefficient, especially if freshness should guaranteed. \textbf{OSP} involves average bandwidth consumption because each request to the OSP server is single and independent, but still requires a connection to the server to get a response.
\textbf{VSL} has bandwidth consumption much higher than \textbf{RIL}, because the set of status information of VCs is much larger than the set of revocation documents.
\textbf{STA} and \textbf{OSA} result in low Verifier-side bandwidth consumption because the status assertions are sent to the Holder.

\textbf{Bandwidth consumption Holder-side}.
\textbf{RIL}, \textbf{VSL}, and \textbf{OSP} result in low Holder-side bandwidth consumption because the revocation information are downloaded by the Verifier.
\textbf{STA} have high bandwidth consumption for the Holder, who downloads periodically assertions.
\textbf{OSA} involves average bandwidth consumption, as the Holder may have to generate specific requests to obtain detailed information from the status of their credential.

\textbf{Untraceability by the Verifier}.
With \textbf{RIL}, we do not have untraceability because the Verifier can download the list and monitor the status of credentials without the Holder's consent since the list is public. 
\textbf{OSP} also does not support untraceability because it allows the Verifier to track the status of a VC without the need for explicit consent, as requests to the Issuer can be always submitted.
\textbf{VSL} does not support untraceability
because the Verifier can track the status of credentials without the Holder's consent.
\textbf{STA} supports untraceability because the Verifier cannot monitor the VC status without the Holder's consent.  
\textbf{OSA} also offers protection against traceability, as the Verifier is not aware of requests done by the Holder to the Issuer. 

\textbf{Time-flexibility}.
The only candidates to support time-flexibility are \textbf{STA} and \textbf{OSA}, because they support untraceability by the verifier. It is easy to see that they support partial time-flexibility because the Holder can maintain past status assertions and exhibit them to the Verifier. Thus, they support time-flexibility only in the past.

\textbf{Untraceability by the Issuer}.
With both \textbf{RIL} and  \textbf{VSL}, no tracking is possible by the Issuer, because the Verifier
downloads the entire lists, thus not identifying a specific Holder.
Conversely, \textbf{OSP} allows the full tracking by the Issuer because the Verifier submits a specific request to the Issuer that identifies a Holder.
\textbf{STA} does not allow tracking because the Issuer sends status assertions to all the Holders.
\textbf{OSA} supports partial protection against tracking, because the Issuer knows when a VC is used, thus could infer further information.

\textbf{Censorship resistance}.
It is easy to see that the methods that operate in the domain of revocation information 
(i.e., \textbf{RIL} and  \textbf{OSP}) cannot allow censorship by the Issuer, because the absence of such an information does not inhibit the presentation of the VC to the Holder.
Although \textbf{STA} and \textbf{VSL} operate on the domain of status information, they cannot allow censorship by the Issuer, because the dishonest behavior of the Issuer would be immediately detected. 
The only method that is not resistant to censorship is then \textbf{OSA}, because the Issuer could deny the response, thus blocking the VC presentation.

\textbf{Content-flexibility}.
In principle, as both revocation information and status assertion can include rich information, all the methods support content-flexibility.

\textbf{Freshness}.
\textbf{RIL} and \textbf{VSL} (at least in principle) support freshness. Indeed, the Verifier can download the revocation information list 
or the verifiable credential status list at the moment of the verification (with some bandwidth price, heavy for \textbf{VSL}), so that also
last-minute revocations are detected.
Instead, the freshness of \textbf{STA} is bounded by the download frequency, which may be daily or less frequent, so it does not always offer full freshness. However, in principle, the lists could be downloaded by the Verifier for each new VC presentation. 
\textbf{OSP} and \textbf{OSA} support freshness because they respond in real-time with the current status of the VC.
 
Observe that, for scalability, the description of the VC status and restriction information could be represented in compact form, by adopting Merkle-Hash-Trees or Bloom Filters. However, in this case, we pay a price in terms of content-flexibility (which we consider crucial, in this paper), that is fully nullified.

From the analysis above, it emerges that only \textbf{STA} supports full untraceability (both by the Verifier and the Issuer). However, it has the following serious drawbacks: (1) huge storing of information Issuer-side (the whole set of status assertions of all the released VC); (2) high bandwidth consumption Holder-side; (3) absence of time-flexibility; (4) limited freshness.
In addition, this method involves continuous communication between the Holder and the Issuer beyond credential issuing, moving away from the model of SSI and complicating the task for the Issuer.

Our method, which adheres to the same model as \textbf{RIL} (i.e., only revocation information is represented),  supports full (encryption-enforced) untraceability (both by the Verifier and the Issuer) such as \textbf{STA}, but overcomes drawbacks (2), (3), and (4) above, as we will see in Section \ref{sec:solution}. 
Moreover, no communication besides the initial credential issuing is required between the Issuer and the Holder.
In addition, our method supports content-flexibility, censorship resistance, and, for what concerns scalability and bandwidth consumption Verifier-side is comparable with
\textbf{RIL} with only a price in terms of computation effort of the sole Issuer (which we can assume to be well-equipped). 
The comparison among all the approaches is reported in Table \ref{tab:comparison}.


\begin{table*}[h]
    \centering
    \renewcommand{\arraystretch}{1.3} 
    \setlength{\tabcolsep}{6pt} 
    \begin{tabular}{|l|c|c|c|c|c|c|c|c|c|}
        \hline
    \textbf{Approach} & \textbf{\makecell{Scalability}} & \textbf{\makecell{Bandwidth \\ (Verifier)}} & \textbf{\makecell{Bandwidth \\ (Holder)}} & \textbf{\makecell{Untraceability \\ (Verifier)}} & \textbf{\makecell{Time \\ Flexibility}} & \textbf{\makecell{Untraceability \\ (Issuer)}} & \textbf{\makecell{Censorship \\ resistance}} & \textbf{\makecell{Content \\ Flexibility}} & \textbf{\makecell{Freshness}}\\  
        \hline
        \makecell[l]{RIL} & \yes & \partialcircle & \yes& \no & \no & \yes & \yes & \yes & \yes\\
        \hline
        \makecell[l]{OSP} & \no & \yes & \yes & \no & \no & \no & \yes & \yes & \yes\\
        \hline
        \makecell[l]{VSL} & \partialcircle & \no & \yes & \no & \no & \yes & \yes & \yes & \yes \\
        \hline
        \makecell[l]{STA} & \no & \yes & \no & \yes & \partialcircle & \yes & \yes & \yes & \no\\
        \hline
        \makecell[l]{OSA} & \no & \yes & \partialcircle & \yes & \partialcircle & \partialcircle & \no & \yes & \yes\\
        \hline
        \textbf{This paper} & \partialcircle & \partialcircle & \yes & \yes & \yes & \yes & \yes & \yes & \yes \\
        \hline
    \end{tabular}
\caption{Comparison of different approaches. $ \protect\yes$: Feature is fully supported. $\protect\no$: Feature is not supported. $\protect\partialcircle$: Feature is partially supported.}
    \label{tab:comparison}
\end{table*}

\section{The Proposed Solution} \label{sec:solution}
In this section, we present our solution.
We assume that the revocation information are richer than standard information. Thus, the revocation of a VC is a document, in any format, claiming a given restriction status. For example, the driver's license point system includes both temporal and spatial restrictions. Depending on the type of offense committed, a driver's license may be revoked permanently, or suspended temporarily, or the driver may be subject to territorial restrictions on its use. 

In our protocol, we consider the following actors.

\begin{itemize}
    
    \item Issuer $I$: who issues the VC and attests to its status.
    
    \item Verifier $V$: who verifies the status of the VC.
    
    \item Holder $H$, who owns the VC.
    
    \item We assume there is a PKI for signatures and certificates verification.

    \item The PKI of a (quantum-safe) Anonymous Hierarchical Identity-Based Encryption (AHIBE). The PKI is a trusted entity (eventually a threshold cryptography-based could be adopted to require the PKI to be semi-trusted). In the case of the EUID wallet, this may be performed by a Supervisory Body (i.e., AGID in Italy).
    
\end{itemize}

Furthermore, our protocol involves the following steps.

\noindent
\textbf{Initial Issuer Set-up}. The Issuer initializes a hash table (assumed based on the compression function \textbf{mod}), by setting the size $d$ on the basis of the expected storage.
$d$ is public. We call this data structure \textit{revocation information table}.

\noindent
\textbf{Initial Holder Set-up}. The PKG gives $H$ the private key associated with their identity.  
    
\noindent
\textbf{Issuing Verifiable credentials}. 
    In our model, a VC is a document in any format that refers to a particular Holder and contains a statement about it.
    We assume that each VC is identified by an ID, say $x$. The VC is signed by the Issuer, and any party can verify its authenticity and integrity by relying on the PKI. 
    This is the phase in which the Issuer releases the VC to the Holder. For each credential, a secret $S_x$ is exchanged, which is the seed of a PRNG.
    Every day, the PRNG is increased. On day $k$ after VC issuing, the Issuer publishes the MAC $MAC(PRNG_k(S_x), x)$, where $h$ is a cryptographic hash function.
    MACs are organized in the form of the hash table, with proper sizing, in such a way that the Verifier has to download only a segment of the hash table to reduce bandwidth consumption while preserving anonymity.
    We call this structure \textit{check table}.
    The domain of the digest is such that reversing is infeasible (i.e., the cost is that of the brute-force attack).
    In this phase, suitable information is set to allow the Holder to exhibit the VC together with the corresponding \textit{proof of possession}.
    
\noindent
\textbf{Publication of Revocation Information}. We suppose that the Issuer on the day $T$ has to publish revocation information $R_k$ that concerns the VC identified by $x$ referred to $H$. 
    Also, we assume that $T$ is the $k$-th day of the VC life. 
    Thus, $I$ computes: $x'= E_{H||T} (MAC(PRNG_k(S_x), x)) \ \textbf{mod} \  d$, where $PRNG_k(S_x)$ is the $k$-th PRNG computed starting from the seed $S_x$.
    Recall that $E_{H||T}$ denotes the encryption for the identity $H||T$, which represents the ephemeral (daily) identity associated with $H$.
    Then, $E_{H||T}(R_x)$ is inserted in the revocation information table in the position $x'$.

\noindent
\textbf{Daily Issuer Set-up}. Let $T'$ be the current day. All the revocation documents of the previous day are re-inserted by using, for each holder $H$, the identity
    $H||T'$. The computation can be performed offline, even over several days, allowing only incremental parts to be integrated.
    Moreover, all the digests $MAC(PRNG_k(S_x), x)$ in the check table are updated with the new PRNGs (i.e., the check table is re-organized). Also this task can be pre-computed in advance.

\noindent
\textbf{VC presentation}. $H$ submits to $V$ a VC identified by $x$. 

    Suppose we are on day $T$. The Holder performs the following tasks:
    \begin{itemize}
        \item $H$ presents the proof of possession.
        \item $H$ sends both $PRNG_k(S_x)$ and the private key corresponding to the identity $H||T$.
        We note that this key can be generated by the Holder from the private key $H$ (this is a feature of AHIBEs), and the Verifier cannot derive other private keys. 
        In this phase, the Holder may provide temporal authorizations other than the current day, by sending the proper keys and PRNGs. 
    \end{itemize}

\noindent
\textbf{Revocation information check}. First, the Verifier checks the validity of both the private key and the PRNG. 
    The private key is trivially checked by encrypting a random number for the identity $H||T$ and decrypting it with the provided private key.
    For the PRNG, the Verifier finds the digest $MAC(PRNG_k(S_x),x)$ 
    by downloading the proper segment of the check table published by the Issuer. 
    If the digest is not found, then the VC presentation is not legitimate and the protocol halts. 
    Conversely, finding the digest proves the correctness of the association between the received PRNG and the identifier $x$. No other party (but the Issuer) can forge this information. 
    At this point, the Verifier downloads the revocation information table, computes $x'= E_{H||T} (MAC(PRNG_k(S_x), x)) \ \textbf{mod} \  d$  (notice that all information to calculate $x'$ is available), accesses $x'$, and tries to decrypt with the received private key all revocation information occurring in the overflow list associated with $x'$. 
    Thus, if a revocation information for the presented VC has been published, then $V$ finds it.
    We note that $d$ can be set in such a way that the overflow lists are small (even 1 element, in the average) so that the computational effort for the Verifier is negligible.

At this point, we are able to demonstrate how our solution meets the features outlined in Section \ref{sec:motivations} and reported in Table \ref{tab:comparison}.

Our method presents limitations of scalability similar to \textbf{RIL}, with some added complexity, because the Issuer has some cryptographic tasks to perform (this is just the price we have to pay).
However, the possibility to pre-compute in advance the encryption of revocations and update only incremental parts 
reduces the load on serves and allows more efficient management. 
Although the Verifier needs to download a considerable amount of information (i.e., the segment of the check table and the revocation information table) to verify both the authenticity of the PRNG and the presence of revocations, 
the bandwidth consumption Verifier-side is not prohibitive. 
Anyway, the bandwidth consumption Verifier-side of \textbf{VSL} is much greater because the size of the whole domain of VCs is much greater than the size of the revocation information domain.
The bandwidth consumption Holder-side is very low in our method, as it is only required to send the daily PRNG value and the AHIBE private key
to the Verifier, only when a VC is presented. 
The Verifier can track a specific revocation only during the day in which it gains access to it, as the minimum granularity of the authorization is one day, than both the PRNG and the private key are valid for one day. Thus, in practice, we can say that untraceability by the Verifier is fulfilled. 
The Issuer is not able to detect if a VC has been presented because the Verifier does not make specific requests to the Issuer but directly downloads the hash tables. Therefore, untraceability by the Issuer is fully achieved.
Concerning time-flexibility, our protocol is really innovative with respect to the state of the art.
Indeed, it allows the Holder to determine when and for how long the Verifier can access the revocation information, by providing the proper PRNGs and private keys.
Interestingly, they can also regard the past.
Concerning censorship resistance, as the Issuer does not have to provide any information to the Holder necessary for the VC presentation,
it cannot act as a censor.
In principle, as in \textbf{RIL}, it could publish fake revocation information, but this public dishonest behavior would be easily detectable and
would pose a serious risk to the Issuer.
Content-flexibility is obtained as the revocation document allows for the inclusion of any information deemed necessary. 
Finally, freshness is fully achieved, because the hash table including the revocation information can be updated in real-time so that the Verifier can
retrieve also last-minute revocation information.

\section{Security Analysis}\label{sect:security}

In this section, we sketch the security analysis of our solution by showing that the attacks enumerated in Section \ref{sec:threat_model} cannot succeed.

\noindent
\textbf{Attack 1}. Recall that, on the day $T$,  the information included in the revocation information table are encrypted for the identity $H||T$. Being the adopted scheme anonymous, no information about the identity can be derived without the private key. Moreover, the entry $x'$ of the revocation information table in which the possible revocation information of the VC $x$ is included, can be computed only if the proper PRNG is known. 
Therefore, it is impossible for the Verifier to find this position on an unauthorized day. Moreover, even if this position is somehow guessed, the Verifier does not own the private key and cannot generate this key from the key received by the Holder. 
Anyway, positions cannot be derived by the check table without the proper PNRG, because MACs are not reversible.

\noindent
\textbf{Attack 2}. The Issuer receives the request from the Verifier only (not from the Holder). The request coming from the Verifier cannot identify a specific VC. Indeed, any segment of the check table includes a sufficiently large number of MACs (and, then referred VCs). Moreover, the download of the revocation information table does not identify any VC.

\noindent
\textbf{Attack 3}. The Holder does not require anything from the Issuer in the phase of VC presentation. Therefore, censorship resistance is achieved by design.

\noindent
\textbf{Attack 4}. The Holder of the VC identified by $x$ cannot send a fake PRNG to the Verifier with the purpose of addressing the Verifier to an entry of the revocation information table in which revocation documents of $x$ are not included. Indeed, to do this, the Holder should be able to compute the MAC $MAC(y, PRNG_k'(S_y))$ of a valid pair. But, $S_y$ is shared only between the legitimate Holder of $y$ and the Issuer, so that the PRNG cannot be computed (actually, not even $k'$ is known). Obviously, the bling guessing of a valid MAC is impossible for the security properties of MACs. Finally, the Holder cannot provide the Verifier with a fake private key, because it would be immediately detected by the Verifier.

\section{Conclusions}
\label{sec:conclusion}
In this paper, we present a new method for VC revocation in an SSI framework that is robust against honest-but-curious Verifiers and Issuers, Issuers acting as censors, and dishonest Holders. To achieve this, we customize the use of anonymous hierarchical identity-based encryption to implement an encryption-enforced, temporal-based access control mechanism, where access is granted to ephemeral identities that include the authorized time.
Compared to the state of the art, our method offers significant advantages and introduces the novel feature of time-flexibility.
As future work, we plan to further refine our solution by exploring possible optimizations, formally proving the security of our approach, and implementing a proof of concept.

\section*{Acknowledgments}
\label{sect:acks}
This work is partially supported by project SERICS (PE00000014) under the MUR National Recovery and Resilience Plan funded by the European Union - NextGenerationEU and partially supported 
Agenzia per la Cybersicurezza Nazionale under the programme for promotion of XL cycle PhD research in cybersecurity – C36E24000080005.
The views expressed are those of the authors and do not represent the funding institutions. The authors are grateful to Vincenzo De Angelis and Sara Lazzaro for their insightful discussions.



\end{document}